\documentclass[conference]{IEEEtran}
\IEEEoverridecommandlockouts

\usepackage{cite}
\usepackage{amsmath,amssymb,amsfonts}
\usepackage{algorithmic}
\usepackage{graphicx}
\usepackage{textcomp}
\usepackage{xcolor}
\def\BibTeX{{\rm B\kern-.05em{\sc i\kern-.025em b}\kern-.08em
    T\kern-.1667em\lower.7ex\hbox{E}\kern-.125emX}}
\begin{document}

\title{Codec training with implicit bandwidth extension for high fidelity speech synthesis}

\author{\IEEEauthorblockN{Thomas Merritt, Rasmus Dall\textsuperscript{*}, Felipe Espic, Vivian Hu, Felix Vaughan, Ger Carney, Marco Marchini, \\Erik Henriksson, Moritz Haller}
 \IEEEauthorblockA{\textit{Spotify} \\
 tmerritt@spotify.com, felipee@spotify.com, vivianh@spotify.com}
 }

\maketitle

\begingroup
\renewcommand{\thefootnote}{\fnsymbol{footnote}}
\footnotetext[1]{Work done whilst working at Spotify.}
\endgroup

\begin{abstract}
Many modern state-of-the-art speech synthesis systems are based on two independently-trained components: 1) a neural codec, 2) an LLM-like model which predicts codec codes given the text to be generated. Whilst effective on lower fidelity speech, to achieve highly natural high fidelity speech we need to scale the approach to output high fidelity waveforms. The standard approach is to provide the target waveform at the input to the codec model during training, relying on the model to learn codes that reconstruct speech well and are predictable by downstream models. Instead, we propose a codec approach which trains on lower fidelity speech at the input and outputs high fidelity speech. This requires the model to implicitly learn the task of bandwidth extension (BWE). 
The proposed approach is demonstrated to outperform the standard approach for speech synthesis and achieves state-of-the-art BWE performance without requiring additional components.
\end{abstract}


\section{Introduction}
Many modern state-of-the-art speech synthesis approaches are built around two primary independently trained components: 1) a neural codec, 2) an LLM-like transformer-based model to predict the codec codes given the text to be generated (e.g., \cite{wang2023neuralcodeclanguagemodels, lyth2024parlertts, defossez2024moshi, kharitonov2023speartts, ye2025llasa}). 
Whilst effective, these models are typically trained on lower fidelity speech waveforms (16kHz or 24kHz). As we look to get closer to generating fully natural high fidelity speech, we need to explore how to train these models to generate higher fidelity speech waveforms. In addition, there is a risk that these two components are not trained in mutually beneficial ways. That is to say that codec models are trained to reconstruct encoded speech with very high accuracy but do not necessarily result in codes that a downstream model will find easy to predict.

The standard approach to adjusting the approach to scale up to higher fidelity waveforms is to provide the codec with the high fidelity waveform at the input during training. This relies on the codec model being able to learn codes that will be easy to predict and that reconstruct well to output high fidelity speech. An alternative is to train a bandwidth extension (BWE) component after the codec. This takes the waveform predicted by the codec at a lower fidelity and adds the missing detail to output high fidelity speech. Whilst effective this adds additional complexity during synthesis system training and also adds extra components during inference-time.

In this paper we investigate how to best train the codec model to output highly natural, high fidelity speech whilst learning codes that the downstream model finds easy to predict. 
We find that encoding lower resolution speech and then relying on the decoder of the codec to decode the learnt codes to high resolution speech results in high quality speech without the need for additional BWE components. 
Specifically conducting this codec training in a two-stage manner gives significantly stronger results. 
This is achieved without the need for additional loss components in the codec training. We also demonstrate that the subsequent codes are easy to predict by the downstream transformer-based model.

\begin{figure*}[t]
  \centering
  \includegraphics[width=\linewidth, trim=20mm 45mm 20mm 45mm, clip]{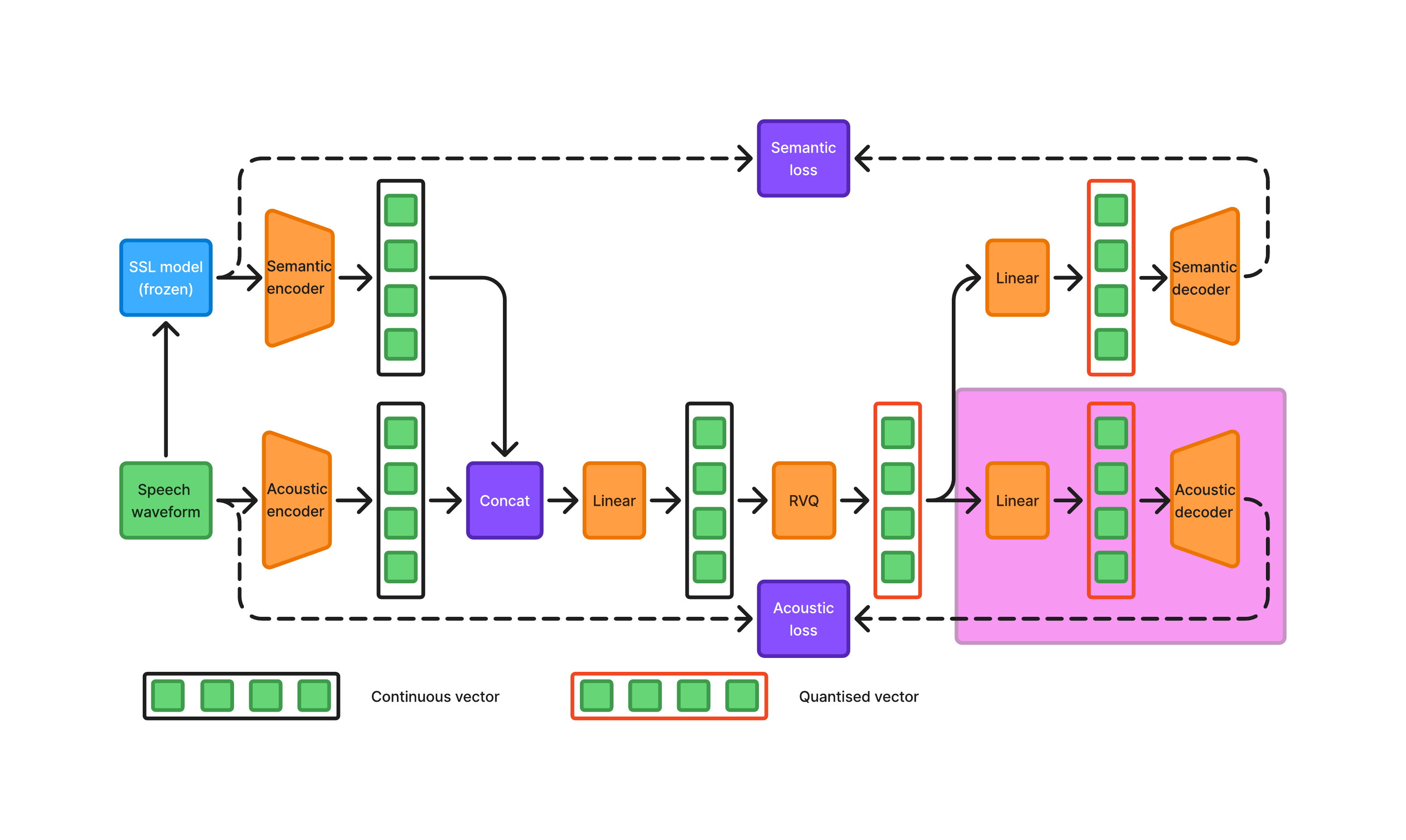}
  \caption{Overview of the X-Codec model. During standard training all learnable weights, shown in orange, are trained. However, during the second stage of the proposed two-stage BWE approach, only the weights shown in the magenta square are trained. All other components are frozen for the second stage.}
  \label{fig:xcodec}
\end{figure*}

\subsection{Previous Work}
\subsubsection{Neural Audio Codecs}
Waveform generation is usually the last step in a TTS pipeline. While vocoders process continuous data representations (e.g., Mel-spectrograms), neural audio codecs operate on discrete codes to generate the synthesised waveform. Vocoders such as MelGan \cite{MelGAN}, Hifi-GAN \cite{hifigan}, or BigVGAN \cite{BigVGAN} have been fundamental to the development of neural audio codecs. Their autoencoder-based structure, training strategies using reconstruction and adversarial losses, and the use of convolutional and transposed convolutional blocks as main building blocks remain standard in audio codecs.  

With the rise of LLMs, neural audio codecs have taken on a significant role as the final stage in the pipeline of SOTA TTS models. Soundstream \cite{SoundStream} based on Hifi-GAN, is a neural audio codec whose encoder, decoder, and quantiser are trained end-to-end. It leverages 
residual vector quantizer (RVQ), and a quantiser dropout technique, which allows the same model to run at different bitrates. EnCodec \cite{EnCodec}, uses a VQ-commitment loss that significantly reduces artefacts along with a small transformer-based language model that acts on the quantised units to further reduce the bandwidth. DAC \cite{dac} is based on Soundstream, and uses Snake activation functions as in BigVGAN. It improves quality and compression ratio over its predecessors by incorporating vector quantisation techniques derived from image processing, along with improvements to the reconstruction and adversarial loss modules.

Despite these improvements, these codecs were originally designed as audio compressors for tasks such as transmission or storage, which may make them suboptimal to work with LLM-based TTS models. SpeechTokenizer \cite{speechtokenizer} was designed with this shortcoming in mind and includes semantic features extracted by a pre-trained semantic encoder. X-Codec \cite{ye2025codec} improves on this idea by using a semantic reconstruction loss after the RVQ and integrates semantic and acoustic features into a unified representation, making it easier for external models such as LLMs to work with. Due to its high performance and speed, X-Codec was chosen as the basis for our work.

\paragraph*{Addendum}
Since this work was initially conducted there have been complementary studies published. 
\cite{huang26l_interspeech} explores training codecs simultaneously with multiple input and output sample rates. In \cite{huang26l_interspeech}, they first pre-trained with the highest sample rate before introducing the range of different sample rates, whereas here we explore different combinations to attempt to output high fidelity speech. In addition, in \cite{huang26l_interspeech} the ability for the model to take speech from lower sample rates and perform bandwidth extension to improve fidelity was not evaluated. Instead only scenarios where input and output sample rates match were used for training and evaluation. Comparison with the approach in \cite{huang26l_interspeech} is required future work. 
In \cite{lemerle26_interspeech} a similar two-stage training methodology to the one proposed here for improving generated speech fidelity was found to perform well for the reduction of the codec frame rate. Note that in \cite{lemerle26_interspeech} the encoder and decoder from the first training stage are fully preserved and the second stage trains components that sit between these. 
\cite{yang26n_interspeech} studies how to best design the codec for highest quality, managing the trade-off between larger number of codes vs larger number of codebooks. In the study presented in this paper we held these hyperparameters fixed across systems evaluated. However, \cite{yang26n_interspeech} highlights the challenge of optimising for the codec and speech synthesis scenarios, as the best performing setup is different for the two scenarios. 
Note that the studies presented in \cite{lemerle26_interspeech} and \cite{yang26n_interspeech} did not focus on high fidelity audio.

\subsubsection{Bandwidth Extension (BWE)}
Here we investigate neural audio codecs decoding speech codes extracted from lower resolution waveforms into high resolution waveforms. That is, the decoder performs bandwidth extension (BWE) in addition to its main task of speech decoding.

Classic BWE methods were carefully designed DSP-based algorithms used mainly for telecommunication applications, e.g., 
based on linear prediction \cite{yasukawa96_icslp_lpc_bwe} and spectral folding \cite{spectral_folding_bwe}. Despite working 
well for their intended tasks, they have become insufficient for the demands of today’s applications, since they were designed to extend to limited bandwidths (e.g., 8kHz).

To mitigate these drawbacks, neural-based BWE methods have been proposed. \cite{Kuleshovetal2017_audiounet} was designed specifically for BWE avoiding any DSP step in the process, directly working in the waveform domain using a UNet-like \cite{unet} architecture. Voicefixer \cite{voicefixer} and AudioSR \cite{AudioSR} aim at improving the efficiency by operating in the spectral domain, using Mel-spectrograms as an intermediate representation, and then using a vocoder to generate the higher resolution waveform. 
\cite{liu22x_interspeech} investigates flexibility in the input resolution whilst performing BWE. 
Instead of using a vocoder as the final stage, AERO \cite{AERO} directly predicts complex spectra, which is converted to the signal domain by an inverse Short-Time Fourier Transform (iSTFT). By doing so, AERO is capable of modelling and predicting phase information, which is otherwise lost when using Mel-spectrograms. 
Whilst AERO implicitly reconstructs amplitude and phase, a newer approach AP-BWE \cite{lu2025towards} explicitly optimises amplitude and phase prediction, each in its own stream, solving the phase-wrapping problem with three anti-wrapping losses and an arctan2 parameterisation. This makes AP-BWE a strong BWE approach to baseline against in this investigation. 

Diffusion-based models have been successful for BWE. NU-Wave \cite{NU-Wave} was the first of this type, combining characteristics of the neural vocoders DiffWave \cite{diffwave} and WaveGrad \cite{wavegrad}. In contrast, UDM+ \cite{UDM+} leverages a noise predictor conditioned by low resolution audio and a novel sampling algorithm that propagates information through the reverse sampling process of diffusion models, which makes it a strong baseline. Even though diffusion-based methods achieve remarkable perceptual audio quality, they suffer from the typical slow inference speed of diffusion models due to the large number of denoising steps required.

Recently there have been some investigations which have explored using different fidelity speech at the input vs the output of the codec, similar to the approach we investigate here. 
In \cite{atamanenko2025tts} and \cite{huang2026kanade} 16kHz audio is encoded at the input with higher fidelity output speech targeted. 
\cite{atamanenko2025tts} investigated 16kHz, 24kHz and 48kHz output speech, whilst \cite{huang2026kanade} only explored generating 24kHz speech, which is lower fidelity than investigated here. 
The approaches taken in both these papers are consistent with the `direct' approach investigated in this paper. However, this was not compared against the `two stage' approach we explore here. 
In \cite{ning2025diffrhythm}, the authors explore the use of mp3 input audio. However, this is only as a form of data augmentation, designed to improve the robustness of the decoder. The encoder weights were not trained on mp3 input audio and the approach does not change the sample rate of the input speech. 
Finally, \cite{liu25p_interspeech} proposed a codec intended for performing both denoising and BWE. It leverages a pre-trained codec, whose codes are cleaned by an external conformer-based deep neural network. However, the investigation focused on bandwidth extension of 8kHz speech up to 16kHz speech, much lower than the fidelity we are targeting here.

\section{Proposed approach}

\subsection{X-Codec}
This investigation focuses on 
learning to implicitly perform bandwidth extension with the X-Codec model \cite{ye2025codec}. 
This approach combines representations extracted from an audio encoder component with a semantic representation extracted from a pre-trained and frozen self-supervised model (WavLM \cite{chen2022wavlm}), as shown in Figure \ref{fig:xcodec}. The representations from these two components are combined 
in the codes with losses ensuring that both acoustic and semantic aspects are preserved. Thus, the encoded audio is reproducible by the acoustic decoder and the semantic vector is reproducible by the semantic decoder. 
X-Codec applies residual vector quantisation (RVQ) with a delay pattern \cite{copet2023simple}. 
A Mean Squared Error (MSE) loss is used for the semantic reconstruction loss. The acoustic reconstruction loss combines multiple components: a generative adversarial network (GAN) loss, a full-band STFT loss and a sub-band STFT loss. 
The full-band STFT loss is a series of STFT losses, comparing generated speech against the target, applied with differing 
FFT sizes, window sizes and hop sizes. 
These STFT losses are then summed together to give the full-band STFT loss. 
The sub-band STFT loss first applies Pseudo Quadrature Mirror Filter (PQMF) to the generated and target speech signals to decompose them into different frequency bands \cite{nguyen1994near}. STFT losses are then applied to this decomposition with 
differing FFT sizes, window sizes and hop sizes. 
These are summed together to produce the sub-band STFT loss. 
A commitment loss is also applied to the quantiser component. The code implementation used in this investigation is based on the 
code shared by the X-Codec authors\footnote{https://github.com/zhenye234/xcodec}. We refer to \cite{ye2025codec} for further details of X-Codec.

\subsection{Bandwidth extension approach}
\label{subsec:bwe_approach}

In this investigation we hypothesise that it is challenging to scale the standard codec approach to higher fidelity sample rates. The original X-Codec paper investigated 16kHz waveforms. High quality speech requires higher sample rates. As the sample rate is increased it increases the amount of high frequency detail present. 
We hypothesise that this additional detail makes it hard for the model to learn robust codes which allow for reconstruction of the waveform and remain predictable by the downstream TTS model. 
Note that we state here that we want the codes to remain predictable rather than improving predictability. This is because we want to avoid overfitting the codec scenario and then experiencing a quality drop when they are applied to the TTS scenario.

Here we investigate instead training the codec model on lower sample rate speech at the input but outputting higher sample rate speech from the acoustic decoder. To achieve this, the acoustic encoder and decoder compression and decompression rates in the codec are adjusted so they are no longer equivalent (the rates used depending on sample rate are shown in Table \ref{tab:rates}). 
We hypothesise that this results in a model which converges 
better as it focuses on learning codes from the fundamental speech structure. 
However, crucially the codec has been trained to perform bandwidth extension (BWE) implicitly in order to fill in the gaps in higher frequencies.  
We investigate two methods to training this: 1) direct training, 2) a two-stage approach. 

\begin{table}[]
\caption{Rates applied based on sample rate. Depending on whether this is being applied during encoding or decoding, these are either downsampling or upsampling rates.}
\centering
\begin{tabular}{lc}
\hline
Input/output & Downsampling/upsampling \\
sample rate & rates applied \\
\hline
8kHz & [5, 4, 4, 2, 2] \\
16kHz & [5, 4, 4, 2, 2, 2] \\
48kHz & [5, 4, 4, 3, 2, 2, 2] \\
\hline
\end{tabular}
\label{tab:rates}
\end{table}

During direct training (similar to \cite{atamanenko2025tts,huang2026kanade}) we adjust the acoustic decoder decompression rates and train the 
model as usual, using all losses. However, for the two-stage approach we first train the codec model on lower sample rate speech following the standard approach (i.e., input and output speech are of the same sample rate). Once the model is converged, we perform a second stage of training where the encoders and RVQ components are frozen, effectively freezing the codes. We replace the codec decoder to output higher sample rate speech. The model is then trained using only the acoustic reconstruction loss (shown in Figure \ref{fig:xcodec}).

\section{Experimental setup}
To test the capability of codecs to implicitly learn BWE and test the predictability hypothesis, we 
evaluate three different BWE system types:
1) Standard 16kHz X-Codec with external BWE approaches from literature applied ($classic\_spectral\_folding\_16kHz{-}48kHz$ \cite{spectral_folding_bwe}, $UDM{+}\_16kHz{-}48kHz$ \cite{UDM+}, $AP{-}BWE\_16kHz{-}48kHz$ \cite{lu2025towards}). 
2) Proposed direct BWE approach as described in Section \ref{subsec:bwe_approach} ($proposed\_16kHz{-}48kHz\_direct$),
3) Proposed two-stage BWE approach as described in Section \ref{subsec:bwe_approach} ($proposed\_16kHz{-}48kHz\_two\_stage$). 
We compare these against training the X-Codec model at 48kHz ($X{-}Codec\_48kHz$) following the standard approach (i.e., where input and output speech are at 48kHz). 
Finally we compare these approaches against upsampling without BWE, using the sinc function ($sinc\_16kHz{-}48kHz$).

We also created 8kHz to 48kHz versions 
to test if further simplifying the 
input speech 
brings further improvements. 
In total we trained 5 versions of X-Codec with 7 differing decoders 
(the BWE approaches from literature and the proposed two-stage BWE approach share the same encoder weights). 
All codec models were trained for 1 million steps with a batch size of 32. 
All codecs were trained to provide the same code sequence lengths (25Hz) and used the same number of codebooks (8), each with the same number of codes (1024). 
All models were trained on the same proprietary dataset of 50k hours of 48kHz English speech audio. The dataset features single-speaker recordings from many speakers. 
Note that since the semantic component (WavLM) in X-Codec is an external frozen model trained on 16kHz audio, this component is always provided with 16kHz audio in these experiments. 
We only vary the sample rate for the acoustic elements of the model.

To check if the codec learns BWE we first 
conducted 
analysis-synthesis 
evaluations. 
30 listeners were presented with the same sentence re-synthesised by each codec and asked to rate the audio quality of each sample on a 0-100 scale in a MUSHRA-like evaluation setup \cite{series2014method}.
Evaluators were prompted to ``Please rate the audio quality of the following audio samples''. 
Each system was presented side-by-side in a random order and the original 48kHz audio waveform was included as a reference. 
Each listener rated a random sample of 20 out of 190 unique sentences from 38 differing speakers, resulting in 600 ratings per system.
We also applied Log-Spectral Distance (LSD) \cite{lsd} to objective evaluations for comparison. To assess the preservation of the low band and the reconstruction quality of the high band, we additionally report LSD-LF and LSD-HF, computed over the frequency ranges below and above the cutoff frequency that separates the low band from the extended high band. The power spectra are computed with FFT\_size = 2048 and hop\_size = 512. We also provide numbers for the pure 48kHz X-Codec model as a comparison point to the BWE models and to measure its preservation of both bands.

Next, we trained TTS systems using 
codes from each of the different codec models as the training targets. 
The TTS system topology follows the teacher model in \cite{henriksson2025knowledge}. This is an LLM-inspired transformer based system broadly similar to that of VALL-E \cite{wang2023neuralcodeclanguagemodels} with approximately 1.2b parameters. Each TTS model was trained for 200k steps with a batch size of 512 and the only varying factor was the target codec codes.

From each TTS system, 51 sentences were synthesised and the naturalness 
was 
rated 
by 30 listeners. 
Listeners were asked ``Please listen to each audio sample fully and then rate the naturalness of the speech in the sample''. 
Each system was presented side-by-side in a random order in a MUSHRA-like setup \cite{series2014method}. Each listener rated 20 unique sentences, resulting in 600 ratings per system. 
We only included sentences for which all systems didn't have stability issues. 
For objective evaluations we render 2000 utterances across 365 different voices. 
To measure 
intelligibility, and therefore how stable the TTS model is when trained on different codec codes, we compute word error rate (WER) using WhisperX \cite{bain2023whisperx}. To measure speaker similarity we compute cosine similarity between a recording from the target speaker and the generated speech using WavLM \cite{chen2022wavlm}.

For analysing all subjective evaluation results, we used the Wilcoxon signed-rank test to check if differences between rated systems are statistically significant. We applied Holm-Bonferroni correction to account for the number of condition pairs being compared. A p-value threshold of 0.05 was applied. 

\section{Results and Discussion}

\begin{table}[]
\caption{Log-Spectral Distance (LSD) objective results for codec analysis-synthesis
}
\centering
\begin{tabular}{lccc}
\hline
Method & LSD & LSD-LF & LSD-HF \\
&& ($<=$ 4kHz) & ($>$ 4kHz) \\
\hline
sinc\_8kHz{-}48kHz & 3.18& 1.09& 3.40\\
UDM+\_8kHz{-}48kHz & 1.85& 1.10& 1.97\\
AP-BWE\_8kHz{-}48kHz & 1.78& 1.12& 1.88\\
proposed\_8kHz{-}48kHz\_direct & 1.69& 1.15& 1.76\\
proposed\_8kHz{-}48kHz\_two\_stage & 1.74& 1.15& 1.82\\
X-Codec\_48kHz & 1.53 & 1.03 & 1.60 \\
\hline 
\hline 
Method & LSD & LSD-LF & LSD-HF \\
&& ($<=$ 8kHz) & ($>$ 8kHz) \\
\hline
sinc\_16kHz{-}48kHz & 2.72& 1.12& 3.17\\
classic\_spectral\_folding\_16kHz{-}48kHz & 1.89& 1.11& 2.13\\
UDM+\_16kHz{-}48kHz & 1.68& 1.13& 1.89\\
AP-BWE\_16kHz{-}48kHz & 1.72& 1.17& 1.93\\
proposed\_16kHz{-}48kHz\_direct & 1.69& 1.18& 1.86\\
proposed\_16kHz{-}48kHz\_two\_stage & 1.83& 1.20& 2.03\\
X-Codec\_48kHz  & 1.53 &	1.04 & 1.70 \\
\hline
\end{tabular}
\label{tab:lsd-results}
\end{table}

\subsection{Analysis-synthesis}
\label{subsec:results_codec}
Table \ref{tab:lsd-results} summarises the results of the objective Log-Spectral Distance (LSD) test on the re-synthesised audio. 
The proposed models perform similarly to BWE methods from literature, in all LSD, LSD-LF, and LSD-HF metrics. 
Whilst the standard X-Codec approach at 48kHz ($X{-}Codec\_48kHz$), returns the lowest LSD measurements. 
We observe that all approaches clearly outperform the no-BWE scenario ($sinc\_16kHz-48kHz$ and $sinc\_8kHz-48kHz$), demonstrating that relevant information is being added into the 
waveforms. 
These findings 
support our hypothesis that the codec can learn BWE implicitly during training. 
Note that for $X{-}Codec\_48kHz$ where the cutoff is at 4kHz, both LSD-LF and LSD-HF are lower than their corresponding values for the cutoff at 8kHz. This is perfectly possible because LSD is essentially a root-mean-square over frequency bins, thus two LSD values cannot be compared directly when they are averaged over a different number of bins. Following these findings we select the best baseline BWE approach for the 16kHz and 8kHz input waveform conditions (namely $UDM{+}\_16kHz{-}48kHz$ and $AP{-}BWE\_8kHz{-}48kHz$) for the remainder of the evaluations.

Results from the codec analysis-synthesis subjective evaluations are shown in Table \ref{tab:16khz_codec_eval}. All differences between systems are found to be statistically significant 
except for between $proposed\_16kHz{-}48kHz\_two\_stage$ and $UDM{+}\_16kHz{-}48kHz$, where there is no significant difference. 
The change in ordering between the 16kHz-based approaches and $X{-}Codec\_48kHz$ compared to the objective results demonstrates that training the codec on 16kHz input audio to learn the codec codes is more effective for perceptually-relevant reconstruction. Note that $X{-}Codec\_48kHz$ and the proposed approaches share the exact same decoder structure. $UDM{+}\_16kHz\_to\_48kHz$ further reinforces that it is preferable to learn 
codec codes on lower sample rate speech. 

\begin{table}[]
\caption{16kHz codec analysis-synthesis subjective evaluations}
    \centering
    \begin{tabular}{lc}
    \hline
    Method & Naturalness \\
    \hline
    proposed\_16kHz{-}48kHz\_direct & 54.96 \\
    proposed\_16kHz{-}48kHz\_two\_stage & 58.41 \\
    X-Codec\_48kHz & 42.87 \\
    UDM+\_16kHz{-}48kHz & 59.11 \\
    recording & 72.96 \\
    \hline
    \end{tabular}
    \label{tab:16khz_codec_eval}
\end{table}

\begin{table}[]
\caption{16kHz codec analysis-synthesis baseline comparison subjective evaluation}
    \centering
    \begin{tabular}{lc}
    \hline
    Method & Naturalness \\
    \hline
    X-Codec\_16kHz & 41.92 \\
    proposed\_16kHz{-}48kHz\_two\_stage & 54.48 \\
    X-Codec\_48kHz & 42.1 \\
    recording & 72.94 \\
    \hline
    \end{tabular}
    \label{tab:16khz_codec_baseline_eval}
\end{table}

To confirm that the evaluated BWE approaches were indeed all perceptual improvements over the standard 16kHz X-Codec without BWE, we run an additional evaluation. The results for this are shown in Table \ref{tab:16khz_codec_baseline_eval}. All differences are statistically significant, except for between $X-Codec\_16kHz$ and $X-Codec\_48kHz$. This demonstrates that the proposed BWE approach is producing significant perceptual improvements over the baseline X-Codec approach.

Next we investigated how further reducing the input sample rate to 8kHz impacts signal quality. Results are shown in Table \ref{tab:8khz_ablation_codec_eval}. All differences between systems are found to be statistically significant. This demonstrates that once the sample rate drops below 16kHz we start to lose information from the input speech such that BWE becomes a much more challenging task. We do, however, find that at this reduced sample rate the proposed approach begins to outperform the state-of-the-art BWE approach ($AP{-}BWE\_8kHz\_to\_48kHz$). Another notable observation is that at 8kHz, the direct training approach is preferred to the two-stage approach, the opposite of the findings at 16kHz.

\begin{table}[]
\caption{8kHz ablation codec analysis-synthesis subjective evaluations}
    \centering
    \begin{tabular}{lc}
    \hline
    Method & Naturalness \\
    \hline
    proposed\_16kHz{-}48kHz\_two\_stage & 66.39 \\
    proposed\_8kHz{-}48kHz\_direct & 65.25 \\
    proposed\_8kHz{-}48kHz\_two\_stage & 62.75 \\
    AP-BWE\_8kHz{-}48kHz & 59.43 \\
    \hline
    \end{tabular}
    \label{tab:8khz_ablation_codec_eval}
\end{table}

\subsection{TTS}
\label{subsec:results_tts}
Results from objective evaluations are shown in Table \ref{tab:tts_objective_eval}. From this we can see that training codecs on 16kHz 
input 
waveforms results in much more stable downstream TTS models. 
$proposed\_8kHz{-}48kHz\_direct$ also achieves similarly good stability, however all other models suffer from reduced stability, resulting in decreased intelligibility. 
In addition, these results show that reducing the sample rate below 16kHz at the input to the codec reduces the level of speaker similarity in generated speech. This 
suggests there are vital indicators of speaker identity 
between 4kHz and 8kHz which the codec is unable to otherwise predict. There is, however, no noticeable drop in speaker similarity between codecs trained on 16kHz input waveforms and those trained on 48kHz.

\begin{table}[]
\caption{TTS objective evaluations. Speaker similarity results are shown 
    ``lower quartile / median / upper quartile''.}
    \centering
    \begin{tabular}{lcc}
    \hline
    Method & WER &  Speaker sim \\ 
    \hline
    X-Codec\_8kHz & 0.26 & 0.94 / 0.96 / 0.97 \\ 
    proposed\_8kHz{-}48kHz\_direct & 0.05 & 0.92 / 0.94 / 0.96 \\ 
    proposed\_8kHz{-}48kHz\_two\_stage & 0.25 & 0.94 / 0.96 / 0.97 \\ 
    \hline
    X-Codec\_16kHz & 0.06 & 0.96 / 0.97 / 0.98 \\ 
    proposed\_16kHz{-}48kHz\_direct & 0.03 & 0.96 / 0.97 / 0.98 \\ 
    proposed\_16kHz{-}48kHz\_two\_stage & 0.06 & 0.96 / 0.97 / 0.98 \\ 
    \hline
    X-Codec\_48kHz & 0.17 & 0.96 / 0.97 / 0.98 \\ 
    \hline
    \end{tabular}
    \label{tab:tts_objective_eval}
\end{table}

Table \ref{tab:16khz_tts_eval} summarises the results of the perceptual test on TTS-synthesised audio. 
All differences between systems are found to be 
statistically significant except for between $proposed\_16kHz\_to\_48kHz\_two\_stage$ and $UDM{+}\_16kHz\_to\_48kHz$. 
These results mirror those of the analysis-synthesis evaluations in Section \ref{subsec:results_codec}. 
The proposed approach again outperforms the standard X-Codec approach at 48kHz. 
Again the proposed two-stage approach outperforms the direct training approach, resulting in state-of-the-art BWE without requiring additional components after the codec. This is despite the model only learning to do BWE implicitly.

Next we investigate how further reducing the input sample rate impacts TTS naturalness. Results are shown in Table \ref{tab:8khz_ablation_tts_eval}. All differences between systems are 
statistically significant. Similarly to Section \ref{subsec:results_codec}, we observe that as the input sample rate to the codec is reduced below 16kHz, information is lost which reduces 
naturalness. 
Again, we observe that the proposed approach applied to 8kHz speech outperforms the state-of-the-art BWE approach. 
For the speech synthesis task we observe that the two-stage approach to the proposed BWE codec training outperforms the direct training approach. This is in contrast to the findings from the analysis-synthesis evaluations. 
From informal listening we find that 
synthesised 
speech from the two-stage approach has better signal quality in addition to improvements in prosody. 
This indicates that whilst capable of higher-quality reconstruction under the analysis-synthesis scenario for this particular input sample rate, 
codes learnt with the direct approach are more challenging for downstream models to predict. 

\begin{table}[]
\caption{16kHz TTS subjective evaluations}
    \centering
    \begin{tabular}{lc}
    \hline
    Method & Naturalness \\
    \hline
    proposed\_16kHz{-}48kHz\_direct & 52.58 \\
    proposed\_16kHz{-}48kHz\_two\_stage & 57.44 \\
    X-Codec\_48kHz & 45.16 \\
    UDM+\_16kHz{-}48kHz & 57.08 \\
    \hline
    \end{tabular}
    \label{tab:16khz_tts_eval}
\end{table}

\begin{table}[]
\caption{8kHz ablation TTS subjective evaluations}
    \centering
    \begin{tabular}{lc}
    \hline
    Method & Naturalness \\
    \hline
    proposed\_16kHz{-}48kHz\_two\_stage & 64.99 \\
    proposed\_8kHz{-}48kHz\_direct & 54.35 \\
    proposed\_8kHz{-}48kHz\_two\_stage & 61.62 \\
    AP-BWE\_8kHz{-}48kHz & 58.59 \\
    \hline
    \end{tabular}
    \label{tab:8khz_ablation_tts_eval}
\end{table}

\section{Conclusion}
In this paper we investigate an implicit bandwidth extension approach applied to X-Codec. The 
investigation demonstrates that training codec codes on lower fidelity speech waveforms (16kHz) at the input and training the codec to perform BWE in its decoder results in significant improvements over the standard training approach. We investigate two approaches to 
this: 1) direct BWE training, 2) a two-stage approach. From the investigation it is clear that the two-stage approach performs best. This allows for the model to learn codes on lower fidelity speech and then optimise the decoder to the task of BWE. The proposed approach demonstrates state-of-the-art BWE performance, without needing additional 
losses for this task and without requiring any additional components after the codec. 
We also investigated whether there are further gains from reducing the input sample rate further to 8kHz, however this introduces significant degradations.

Whilst we have robustly investigated the proposed approach across both analysis-synthesis and TTS tasks, this investigation has focused on the X-Codec approach and English language data. As stated in the introduction, we believe that this is a good representation of current state-of-the-art codecs. However, future work is required to confirm that these findings are replicated across other codecs and languages.

\section{Ethical considerations}
We recognise that high-fidelity speech synthesis systems capable of closely replicating a speaker’s voice carry a risk of misuse, such as unauthorised voice cloning and impersonation. All training and evaluation data is proprietary and was used with the speakers' consent for speech synthesis research. While we publish our approach, we do not release the trained models, limiting the risk of misuse.

\bibliographystyle{IEEEtran}
\bibliography{IEEEabrv,mybib}

\end{document}